\documentclass[conference]{IEEEtran}
\IEEEoverridecommandlockouts
\usepackage{cite}
\usepackage{amsmath,amssymb,amsfonts}
\usepackage{algorithmic}
\usepackage{graphicx}
\usepackage{textcomp}
\usepackage{xcolor}
\DeclareMathOperator*{\argmin}{arg\,min}

\newtheorem{lemma}{Lemma}

\newtheorem{remark}{Remark}
\newtheorem{assumption}{Assumption}
\newtheorem{example}{Example}
\newtheorem{definition}{Definition}
\newtheorem{proposition}{Proposition}

\def\BibTeX{{\rm B\kern-.05em{\sc i\kern-.025em b}\kern-.08em
    T\kern-.1667em\lower.7ex\hbox{E}\kern-.125emX}}
\begin{document}

\title{Network Design for Social Welfare\\
\thanks{We thank NSF, ONR and The Ripple Foundation for their support.}
}

\author{\IEEEauthorblockN{Abhishek Shende}
\IEEEauthorblockA{\textit{Head of Engineering} \\
\textit{BrilliantMD, Inc.}\\
Austin, TX \\
abhishek@brilliant.md}
\and
\IEEEauthorblockN{Deepanshu Vasal}
\IEEEauthorblockA{
\textit{Department of ECE}\\ 
\textit{Northwestern University}\\
Evanston, IL \\
dvasal@umich.edu}
\and
\IEEEauthorblockN{Sriram Vishwanath}
\IEEEauthorblockA{\textit{Department of ECE} \\
\textit{University of Texas}\\
Austin, TX \\
sriram@utexas.edu}
}

\maketitle

\begin{abstract}
In this paper, we consider the problem of network design on network games. We study the conditions on the adjacency matrix of the underlying network to design a game such that the Nash equilibrium coincides with the objective of the designer and consider the special case when the designer's objective is social welfare. We provide the examples for linear quadratic games that satisfy this condition. Furthermore, we identify conditions on properties of adjacency matrix that provide a unique solution using variational inequality formulation, and verify the robustness and continuity of the social cost under perturbations of the network. Finally we comment on individual rationality and extension of our results to large random networked games. 
\end{abstract}


\section{Introduction}
The surging growth of connections between people in various settings like economic, social and commercial has motivated the studies of how interactions between linked members affect the outcomes. These outcomes not only affect the individuals but also the society, e.g - in terms of total resources consumed, the total profit (or cost). The investigation into underlying networks that promote social good for rational players requires understanding which network structures are beneficial over others to achieve a common community benefit. 

Game theory is a powerful tool to model strategic interactions among competing agents. An important engineering application of game theory is mechanism design, which aims to design systems such that, when played by strategic agents who optimize their individual objectives, they achieve the same objective as envisioned by the designer. Mechanism design can be viewed as reverse engineering of games, where the rules of a game are modified to achieve a desired outcome. The balance between maximization of desired outcome or social goal and other desirable properties such as ease of design, robustness, computational tractability and feasibility is studied in understanding the theoretical underpinnings of mechanism design. A classic and one of the most widely used practical examples of mechanism design is auctions~\cite{Myerson1981}, where an auctioneer requests bids from bidders on a private good. This auction is designed to ensure that,  when these strategic bidders bid on the value to maximize their own valuations, it also maximizes the returns of the auctioneer.  Mechanism design has found use in many important applications; e.g. in distributed scheduling problems, and in public goods where decisions maximize gross social benefit minus the cost of providing the goods. There is rapidly growing literature on the theory of Mechanism Design as well as its real world applications~\cite{Jackson2014}.

Information design, when compared with mechanism design, is a relatively new field introduced by Kamenica and Gentzkow in \cite{Kamenica2011}. This domain studies the manner in which an {\em information designer}, by controlling the information provided to players, can influence their strategic behavior to achieve a desired outcome.
A unified perspective on the basic issues and insights in information design is provided in \cite{Bergemann2019}.

Network games are employed in the study of social and economic interactions, wherein the action of individual player is affected by the actions of its neighbors. Network games have been studied across multiple applications in
\cite{Allouch2015},\cite{Naghizadeh2018},
\cite{Calvo-Armengol2009} 
. Games  on networks have been surveyed in greater detail in \cite{Jackson2015},\cite{Bramoulle2016}. The general focus of literature in the domain of network games has been on understanding the effect of properties of network on  Nash equilibria. Work on network games with strategic complements relate properties of equilibria to the spectral radius of the adjacency matrix of the network in \cite{Jackson2015},\cite{Acemoglu2016},\cite{Ballester2006}, while games with strategic substitutes use minimum eigenvalue of adjacency matrix in \cite{Allouch2015},\cite{Bramoulle2014}. One of the most studied examples in network games are scalar linear quadratic games for which best response is a (truncated) linear function. The equilibria properties for these games are understanding using potential functions in \cite{Monderer1996},\cite{Rosen1965}. Further, these results have been extended using a variational inequality framework in \cite{Melo2018Variational},\cite{Naghizadeh2017}, \cite{Parise2019}.

In this paper, we study the following problem setting:  The designer must enable interactions among the strategic users such that,  presuming strategic players play selfishly on a particular network, the Nash equilibrium of the resulting game must coincide with the designer's objective. Moreover, this Nash equilibrium must correspond to a certain objective.
Analogous to the theory of mechanism design and of information design, to the best of our knowledge, this is the first attempt to analyse \emph{network design} in a manner where selfish players' actions coincide with the designer's objective.  We apply this notion to the case of {\em social welfare}  objective to provide our results.

In the context of scalar linear quadratic games, we derive  conditions on the adjacency matrix under which a solution for the problem exists, i.e.,  where the Nash equilibrium matches social optimum. We use the variational inequality framework to devise a sufficient condition for the uniqueness of such a solution. Further, we characterize the robustness and continuity of the  objective to perturbations in the adjacency matrix and we verify that the individual rationality constraints are satisfied. Finally, we expand this scalar linear quadratic game to a larger network and understand when solutions may exist for other, less structured settings.


\section{Model and Definitions} \label{model_and_def}
A network game $\mathcal{G}$ with set of $N$ players, is played over a weighted directed network whose structure is captured by an $n\times n$ adjacency matrix $G$. The $(i,j)$th entry of $G$, denoted by $g_{ij}$, represents the strength and type of influence of player $j$'s strategy on the cost function of player $i$. The positive(negative) $g_{ij}$ represent strategic substitutes(complements) where an increase in neighbor $j$'s actions leads to a corresponding decrease(increase) in player $i$'s action. We assume $g_{ii} = 0$ for all $i$ $\in$ $\mathbb{N}[1,N]$. The action of a player $i$ is given by $x_i \in \mathcal{X}_i \subseteq \mathbb{R}_{\geq0}$, and so $x = (x_1,x_2\dots x_n) \in \mathcal{X} = \prod_{i=1}^N \mathcal{X}_i$. 
Each player $i$ $\in$ $\mathbb{N}[1,N]$ chooses their action $x_i \in \mathbb{R}_{\geq0}$ to minimize a cost function:
\[J_i(x_i,z_i(x)),\]
which in turn depends on their own action $x_i$ and on the aggregate  neighbors' strategies $z_i(x)$, defined by the weighted linear combination 
\[z_i(x) = \sum_{j=1}^N g_{ij}x_j.\]
The best response for player $i$, i.e., the action that minimizes the cost function is defined as
\[B_i(z_i(x)) :=   \argmin_{x_i} J_i(x_i,z_i(x)).\]
The set of actions within which no player has an incentive for unilateral deviations (i.e., each player is playing a best response to other player's actions) is a Nash equilibrium. Mathematically, a vector $x^{*}  = (x^*_1, \dots x^*_n)$,  is a Nash equilibrium (NE) if, for all players $i$ $\in$ $\mathbb{N}[1,N]$, $x^*_i$ $\in$ $B_i(z_i(x))$.

\subsection{Linear Quadratic Network Game} 

A linear quadratic (LQ) network game is one where each agent chooses a scalar strategy $x_i \geq 0$ in order to minimize the linear quadratic cost function:

\begin{equation} \label{eq:1}
J_i(x_i,z_i(x))= \frac{1}{2}(x_i)^2 +[z_i(x) - a_i]x_i.
\end{equation}

with $a_i$ $\in$ $\mathbb{R}$. Typically, this cost function is parametrized by $K$ which represents the effect of $z_i(x)$ on the equilibrium action of player $i$. However, in our model, this parameter is absorbed within the $G$ matrix.

\section{Nash Equilibrium and Designer's Objective}
The Nash equilibrium (NE) for an LQ game can be derived from the first-order necessary condition to minimize the cost function for each player $i$ given by:
\begin{equation}\label{eq:2}
\frac{\partial J_i(x_i,z_i(x))}{\partial x_i} = x_i - a_i + \sum_{j=1}^N g_{ij}x_j = 0.
\end{equation}
This results in
\begin{equation} \label{eq:3}
    x^*_i = a_i - \sum_{j=1}^N g_{ij}x^*_j.    
\end{equation}
In matrix form 
\begin{equation} \label{eq:4}
    (I+G)x^* = a,
\end{equation}
  
where $I$ is $N\times N$ identity matrix and the vector $a= (a_1, \dots a_n) $.

Based on the their objective, the designer can pick an objective function and design the game so that the optimal point of the objective function matches with the Nash Equilibrium derived in \eqref{eq:4}. We explore a special case of {\em social optimum}, wherein the actions of players  minimizes the sum of individual social cost functions. The conditions derived below for social optimum can be derived similarly for variety of other functions. The social cost function for an LQ game is given by 
\begin{equation} \label{eq:5}
C(x_i,z_i(x)) = \sum_{i=1}^N J_i(x_i,z_i(x))     
\end{equation}

Therefore, a social optimum strategy $y^*_i\geq0$, where the actions of players minimize this social cost would be given by
\begin{equation} \label{eq:6}
\argmin_{x_i} C(x_i,z_i(x)) = \argmin_{x} \sum_{i=1}^N J_i(x_i,z_i(x))     
\end{equation}
\begin{equation}\label{eq:7}
=\argmin_{x} \left\{\sum_{i=1}^N(\frac{1}{2}(x_i)^2 -a_ix_i)+\sum_{i=1}^N\sum_{j=1}^Ng_{ij}x_jx_i\right\}.     
\end{equation}

First order conditions for each $i$ result in 
\begin{equation}\label{eq:8}
y^*_i = a_i - \sum_{j=1}^N g_{ij}y^*_j - \sum_{j=1}^N g_{ji}y^*_j.    
\end{equation}

This, in matrix form, is recast as
\begin{equation} \label{eq:9}
(I+G+G^T)y^* = a.     
\end{equation}
where $y^*= (y^*_1, \dots y^*_n) $.

As desired in our work, the NE and Social optimum coincide when $x^* = y^*$. From \eqref{eq:4} and \eqref{eq:9},
\begin{equation} \label{eq:10}
(I+G)x^* = a,(I+G+G^T)x^* = a,
\end{equation}
which reduces to
\begin{equation} \label{eq:11}
(I+G)x^* = a, G^Tx^* =0.    
\end{equation}

\begin{remark}\label{remark:1}
From \eqref{eq:11}, we observe that, for non-zero $x^*$, $G^T$ is not invertible. Thus, Determinant$(G) = 0$ is a necessary condition for a non-zero NE to coincide with the social optimum. This also implies that at least one of the eigenvalues of $G$ is 0.
\end{remark} 

\begin{remark}\label{remark:2}
If $G$ is symmetric, from \eqref{eq:11}, we observe that  $x^*= a, Ga=0 $. Thus, the NE and the social optimum do not depend on the adjacency matrix and consequently, the payoffs for each player and the social costs are fixed,  depending only on $a$.
\end{remark}

\begin{remark}\label{remark:3}
By \cite{Monderer1996}, a necessary condition for existence of an exact potential function is 
\[\frac{\partial J_i(x_i,z_i(x))}{\partial x_ix_j} = \frac{\partial J_j(x_j,z_j(x))}{\partial x_jx_i}\]
For LQ game, this becomes  $G_{ij} = G_{ji}$. This allows use of theory of potential games for symmetric matrices in socially optimum NE. 
\end{remark}



\subsection{Numerical Examples for N=2,3,4}
\begin{example}
For a two player LQ game, let $G = \begin{bmatrix}0 & g_{12}\\g_{21} & 0\\\end{bmatrix} $. Using \eqref{eq:11},  potential solutions are found only when at least one of $g_{12}$ or $g_{21}$ is zero. 

\end{example}


\begin{example}
For a three player LQ game, 
if $G$ is symmetric, i.e., if $G = \begin{bmatrix}0 & g_{a} & g_{b} \\g_{a} & 0 & g_{c} \\g_{b} & g_{c} & 0 \\\end{bmatrix} $, potential solutions exist only when at least one of $g_a$, $g_b$, $g_c$ is zero.

If $G$ is not symmetric, let $G = \begin{bmatrix}0 & g_{12} & g_{13} \\g_{21} & 0 & g_{23} \\g_{31} & g_{32} & 0 \\\end{bmatrix} $. A numerical example that satisfies \eqref{eq:11} for such $G$ is presented in Figure~\ref{fig:1}.
If $\begin{bmatrix}a_1\\a_2\\a_3\\\end{bmatrix}=\begin{bmatrix}1\\2\\3\\\end{bmatrix}$, and $g_{12}=-2,g_{31}=-3,g_{23}=2 $, using \eqref{eq:11} gives  $g_{21}\approx1.18042 ,g_{13}\approx-0.273107 ,g_{32}\approx37.229, \boldsymbol{x^*} \approx \begin{bmatrix}1.4046\\0.19173\\0.07544\\\end{bmatrix}$.

\begin{figure}[htbp]
\centerline{\includegraphics[width=0.50\columnwidth]{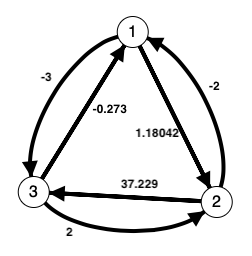}}
\caption{A graph with 3 nodes where Nash equilibrium of the corresponding LQ game coincides with the social optimum solution.}
\label{fig:1}
\end{figure}

\end{example}

\begin{example}
For $N\geq 4$, solutions exist for both symmetric and non-symmetric $G$ where none of the connections between the players are zero. A numerical example for a game with 4 players is presented in Figure \ref{fig:2}.

\begin{figure}[htbp]
\centerline{\includegraphics[width=0.5\columnwidth]{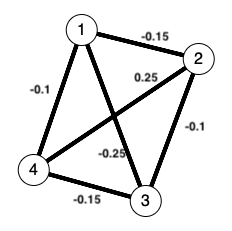}}
\caption{A symmetric graph with 4 nodes and $a_i=1,\forall i$ where Nash equilibrium of the corresponding LQ game coincides with the social optimum solution.}
\label{fig:2}
\end{figure}
\end{example}
\section{Example with non linear cost functions}
In this section, we consider an example where we have a non-linear cost function. Specifically, we consider a modified version of
public goods game from \cite{Melo2018Variational} where
\[J_i(x_i,z_i(x),\theta_i)= \frac{1}{2}(x_i)^2 + [z_i(x)-\gamma_i(\theta_i+z_i(x))]x_i.\]


Here, the function $\gamma_i(.)$ is the demand function associated with public good. The parameter $\theta_i$ may be interpreted as player $i$'s income. This is similar to the LQ game in \eqref{eq:1}, except that the constant $a_i$ is now a function $\gamma_i(\theta_i+z_i(x))$.

The NE $x^*$ from first order condition is given in matrix form as
\begin{equation}\label{eq:12}
(I+G)x^* = \gamma(\theta,z(x^*)).
\end{equation}

The associated social cost is given by 
\begin{equation}\label{eq:13}
    C(x,z(x),\theta) = \sum_{i=1}^N \frac{1}{2}(x_i)^2 +z_i(x)x_i - \gamma_ix_i.
\end{equation}
The social optimum $y^*$ derived from the first order condition 
\[\argmin_{x} \left\{\sum_{i=1}^N(\frac{1}{2}(x_i)^2 -\gamma_ix_i)+\sum_{i=1}^N\sum_{j=1}^Ng_{ij}x_jx_i\right\}\]
is given by
\[y^*_i = \gamma_i(\theta_i+z_i(y^*)) + \sum_{j=1}^Ny^*_j\gamma^{\prime}_j(\theta_j+z_i(y^*))g_{ji} - \sum_{j=1}^N g_{ij}y^*_j - \sum_{j=1}^N g_{ji}y^*_j \]
\begin{equation} \label{eq:14}
  (I+VG^T+G)y^* = \gamma_i(\theta_i+z_i(y^*)), 
\end{equation}
where $V = Diag(1-\gamma^{\prime}_j(\theta_j+z_i(y^*)))$. 

Using $x^* = y^*$, from \eqref{eq:12} and \eqref{eq:14}, the conditions where NE matches with social optimum are

\begin{equation} \label{eq:15}
(I+G)x^* = \gamma(\theta,z(x^*)), (VG^T)x^* = 0.   
\end{equation}

\section{Variational Inequality (VI) Formulation}
Variational Inequality theory provides a general mathematical framework to study solution properties of optimization problems. It has been shown in \cite{Melo2018Variational},\cite{Naghizadeh2017},~\cite{Parise2019} that Variational Inequality framework can be used to characterise the Nash Equilibrium of a network game.
\begin{assumption}\label{assumption:1}
The set $\mathcal{X}_i$ is non empty, closed and convex for all $i \in \mathbb{N}[1,N]$
The cost function $J_i(x_i,z_i(x))$ is continuously twice differentiable with bounded derivatives and convex in $x_i$ for all $i \in \mathbb{N}[1,N]$. Also, $\nabla_{x_i}J_i(x_i,z_i(x))$ is Lipschitz. 
\end{assumption}

\begin{definition}\label{definition:1}
A variational inequality $VI(K,F)$ consists of a set $K \subseteq \mathbb{R}^n$ and a mapping $F : K \rightarrow \mathbb{R}^n$, and is the problem of finding a vector $x^*$, such that 
\begin{equation} \label{eq:16}
F(x^*)^T(x-x^*) \geq 0, \forall x \in K    
\end{equation}
\end{definition}

When Assumption~\ref{assumption:1} holds, a vector of actions $x^*$ is Nash Equilibrium of a network game $\mathcal{G}$ if and only if it solves the $VI(\mathbb{R}_{\geq0}^N, F(x))$ condition with \[F_i(x) = \nabla_{x_i}J_i(x_i,z_i(x))\] 
Similarly, the social optimum is a solution of $VI(\mathbb{R}_{\geq0}^N, F(x))$ condition where \[W_i(x) = [\nabla_{x_i} C(x,z(x))] = \nabla_{x_i} \sum_{i=1}^{N}J_i(x_i,z_i(x))\] and Assumption~\ref{assumption:1} holds for $C(x,z(x))$.

The conditions for existence and uniqueness of Nash equilibria for network games using VI framework has been presented for specific family of best response function \cite{Naghizadeh2017} and extended to more general network games \cite{Parise2019}. We use this approach to derive conditions on uniqueness of socially optimum Nash equilibrium .
We assume Assumption \ref{assumption:1} holds for subsequent analysis.

\subsection{Uniqueness of Socially Optimum NE}
\begin{proposition}\label{proposition:1}
The socially optimum NE on networks satisfying \eqref{eq:11} is unique for LQ game if $ 2-3||G||_2 > 0 $, where $||G||_2$ is the spectral norm. 
\end{proposition}

\textit{Proof.}
Using Definition~\ref{definition:1}, the NE $x^*$ solves \[F(x^*)^T(x-x^*) \geq 0, \forall x\]. The social optimum $y^*$ solves \[W(y^*)^T(x-y^*)\geq 0, \forall x\]. Since $x^*$ = $y^*$
\begin{equation} \label{eq:17}
    F(x^*)^T(x-x^*)\geq 0, W(x^*)^T(x-x^*)\geq 0, \forall x \in \mathbb{R}_{\geq0}^N
\end{equation}
\begin{equation}\label{eq:18}
    (F(x^*)+W(x^*))^T(x-x^*)\geq 0
\end{equation}

From \cite{Facchinei2009}, if $F+W$ is strongly monotone, then the VI has unique solution.
From \cite{Facchinei2007}, $F+W$ is strongly monotone if and only if there exists $\alpha > 0$ such that 
\begin{equation} \label{eq:19}
\frac{\nabla_x(F+W)+\nabla_x(F+W)^T}{2} \succeq \alpha I    
\end{equation}


For network game, $\nabla_{x}F = D_F + K_F$ and $\nabla_{x}W= D_W+K_W$, where $D_F$ and $D_K$ indicates the diagonal matrices consisting of $\nabla_{x_i,x_i}^2$ and $K_D$ and $K_F$ is the matrix with off-diagonal terms that are cross derivatives $\nabla_{x_i,x_j}^2 $. Let $d_F = \min_{x}(\lambda_{min} D_F)$, $d_W = \min_{x} (\lambda_{min} D_W)$.

\begin{align}\label{eq:20}
  &\frac{\nabla_x(F+F^T)+\nabla_x(W+W)^T}{2} \succeq \nonumber\\
 &(d_F +d_W)I + \frac{(K_F+K_F^T)}{2} +\frac{(K_W+K_W^T)}{2} 
\end{align}
\begin{lemma}\label{lemma:1}
For a symmetric matrix $A \in \mathbb{R}^{d \times d}$, $A=A^T$, then following holds,
\begin{enumerate}
\item $A \succeq \lambda_{min}(A)I_d$
\item $|\lambda_{min}(A)|\leq \rho(A) = ||A||_2$
\item $\lambda_{min}(\alpha I+A) = \alpha + \lambda_{min}(A)$
\end{enumerate}
where $||A||_2 = \sqrt{\lambda_{max}(A^TA)}$, $\rho(A)$ is the spectral radius of $A$
\end{lemma}
Using Lemma \ref{lemma:1}.1
\begin{align} \label{eq:21}
     &(d_F +d_W)I + \frac{(K_F+K_F^T)}{2} +\frac{(K_W+K_W^T)}{2} \succeq \nonumber\\
     &\lambda_{min}\left((d_F +d_W)I + \frac{(K_F+K_F^T)}{2} +\frac{(K_W+K_W^T)}{2}\right)I
\end{align}
Using Lemma \ref{lemma:1}.3
\begin{align} \label{eq:22}
   &\lambda_{min}\left((d_F +d_W)I + \frac{(K_F+K_F^T)}{2} +\frac{(K_W+K_W^T)}{2}\right) \geq \nonumber\\
    & (d_F +d_W) +\lambda_{min}\left(\frac{(K_F+K_F^T)}{2} +\frac{(K_W+K_W^T)}{2}\right)
\end{align}

For LQ game, \cite{Parise2019}, shows $D_F=I$,$K_F = G$. Deriving $D_W$ and $K_W$
\[\nabla_{x_i}W(x,z(x)) = x_i-a_i+\sum_jg_{ij}x_j+\sum_j g_{ji}x_j\]
\[\nabla_{x_i,x_i}^2W(x,z(x)) = 1\]
\[\nabla_{x_i,x_j}^2W(x,z(x)) = g_{ij}+g_{ji}\]

Thus, $d_F = d_W = 1$, $K_F = G$ and $K_W=G+G^T$
\begin{align} \label{eq:23}
    \lambda_{min}(\frac{(K_F+K_F^T)}{2} +\frac{(K_W+K_W^T)}{2}) \geq \nonumber\\
     \lambda_{min}(\frac{(G+G^T)}{2} + (G+G^T))
\end{align}
\begin{align} \label{eq:24}
    \lambda_{min}(\frac{(K_F+K_F^T)}{2} +\frac{(K_W+K_W^T)}{2}) \geq \nonumber\\
      \lambda_{min}(\frac{3(G+G^T)}{2})
\end{align}

From  Using Lemma \ref{lemma:1}.2,
\[\lambda_{min}((G+G^T)) \geq -\rho((G+G^T)) = -||(G+G^T)||_2 \]
using the triangular inequality and for any matrix $A$, $||A||_2=||A^T||_2$,
\[\lambda_{min}(G+G^T) \geq -(||G||_2+||G^T||_2) = -2||G||_2\]

Using this result in \eqref{eq:24}
\begin{equation} \label{eq:25}
\lambda_{min}(\frac{(K_F+K_F^T)}{2} +\frac{(K_W+K_W^T)}{2}) \geq -3||G||_2    
\end{equation}

Combining \eqref{eq:20}, \eqref{eq:21}, \eqref{eq:22}, \eqref{eq:25},

\[\frac{\nabla_x(F+W)+\nabla_x(F+W)^T}{2} \succeq (2-3||G||_2)I\]
Comparing \eqref{eq:19}, $ \alpha = 2-3||G||_2 > 0 $ will ensure that social optimum NE from \eqref{eq:11} is unique. 

\begin{proposition}\label{proposition:2}
The socially optimum NE derived from \eqref{eq:11} is unique for LQ game when  $ 2 - (2||G||_{\infty}+||G||_{1}) > 0 $, where $||G||_\infty$ is the maximum absolute row sum and $||G||_{1}$ is the maximum absolute column sum of the matrix .
\end{proposition}

\textit{Proof.}
Variational Inequality formulation $VI(K,F)$ admits an unique solution when F is an uniform block P function \cite{Parise2019}. We define following variables ,
\[\zeta_i^{min} = \inf_{x \in K} \lambda_{min} \nabla_{i}F_i(x), \forall i\]
\[\zeta_{ij}^{max} = \sup_{x \in K} ||\nabla_{j}F_i(x)||, \forall i,j \neq i\]
Using these variables, matrix $\Gamma$ is defined as
\[\Gamma = \begin{bmatrix} 
    \zeta_1^{min} & -\zeta_{i2}^{max} & \dots & -\zeta_{1N}^{max}\\
    -\zeta_{21}^{max} & \zeta_2^{min} & \dots \\
    \vdots & \vdots & \ddots & \vdots\\
    -\zeta_{N1}^{max} &  \dots &\dots & \zeta_N^{min}\\
    \end{bmatrix}\]
For LQ games, using VI formulation in \eqref{eq:18} 
\[\zeta_i^{min} = 2, \forall i\]
\[\zeta_{ij}^{max} = |2g_{ij} + g{ji}|, \forall i,j \neq i\]
Now, \[\Gamma = \begin{bmatrix} 
    2 & -|2g_{12}+g_{21}| & \dots & -|2g_{1N}+g_{N1}|\\
    -|2g_{21}+g_{12}| & 2 & \dots \\
    \vdots & \vdots & \ddots & \vdots\\
    -|2g_{N1}+g_{1N}| &  \dots &\dots & 2\\
    \end{bmatrix}\]
    
\begin{definition}\label{definition:2}
P Matrix - The determinants of all its principal minors is strictly positive.
\end{definition}    
The authors in \cite{Parise2019} show that when $\Gamma$ is P matrix, (F+W) is an uniform block P function, thus providing uniqueness to the solution. Let $M:=2I-g_{max}J$, where $g_{max}$ = maximum element in $|2G+G^T|$ and $J$ is $N \times N$ matrix of ones. For each row $i$ of $M$, 
$\sum_{j\neq i}^N|M_{ij}|\leq ||2G+G^T||_{\infty}$. By Gershgorin circle theorem, every eigenvalue of M is within a circle of centre (2,0) and radius $|2G+G^T|_{\infty}$. 
If 
\[2-||2G+G^T||_{\infty} > 0\]
then every eigenvalue has positive real part, which ensures that determinant (product of eigenvalues) is positive. Similar argument can be applied to principal minors of $M$ and their eigenvalues and determinants. This ensures $M$ is a P matrix by Definition \ref{definition:2}.
As both $M$ and $\Gamma$ are Z matrices (all non diagonal terms are non positive) and $\Gamma \geq M$ element-wise, when $M$ is P matrix leads to $\Gamma$ to be a P matrix.
\[2-||2G+G^T||_{\infty} > 0 \implies 2 - (2||G||_{\infty}+||G^T||_{\infty}) > 0 \]

This $ 2 - (2||G||_{\infty}+||G||_{1}) > 0 $ will ensure that common solution to NE and social optimum from \eqref{eq:11} is unique.

\subsection{Robustness and Continuity}
In this section, we verify the continuity of social cost function for a network where common solution to Nash equilibrium and social optimum exist, under perturbations to the network. 

Under Assumption \ref{assumption:1}, authors in  \cite{Melo2018Variational} have shown that with strong monotonicity of $F(x)$, the Nash Equilibrium $x^*$ varies continuously as a function of $G$. Let the function $F(x,\Bar{G})$ as in \eqref{eq:17} be strongly monotone so that $x^*(\Bar{G})$ is unique Nash Equilibrium for network $\Bar{G}$. Then for any perturbed network $G$ and any NE $x^*(G)$ of $G$,with bounded $\mathcal{X}$ it holds
\[||x^*(G)-x^*(\Bar{G})||_2 \leq \mu (||\Bar{F}(x,\Bar{G})-F(x,G)||_2 \] 
\begin{equation}\label{eq:26}
    ||x^*(G)-x^*(\Bar{G})||_2 \leq \eta (||\Bar{G}-G)||_2\Delta
\end{equation}
where $\Delta=\max_{x\in \mathcal{X}}||x||_2$, $\eta,\mu$ are constants.

The authors in \cite{Parise2019} prove the Lipschitz continuity for Nash equilibrium  when $F(x)$ is a uniform block P function, which is a weaker condition on $F(x)$ than strong monotonicity.   

Using Assumption \ref{assumption:1} and \eqref{eq:5}, the social cost function $C(x,z(x))$ also satisfies Assumption \ref{assumption:1}. 
Thus, using Lipschitz continuity of $C$ with constant $L$,
\begin{align}\label{eq:27}
   & ||C(x^*(G),z(x))-C(x^*(\Bar{G}),\Bar{z}(x))||_2^2 \leq \nonumber\\
    & L^2 (||x^*(G)-x^*(\Bar{G})||_2^2 + ||z(x)-\Bar{z}(x)||_2^2 )  
\end{align}
    
Using \eqref{eq:26} and \eqref{eq:27} , and $z_i(x) = \sum_{j=1}^N g_{ij}x_j$
\begin{equation} \label{eq:28}
    ||C(x^*(G),z(x))-C(x^*(\Bar{G}),\Bar{z}(x))||_2 \leq k (||\Bar{G}-G)||_2\Delta
\end{equation} where k is a constant

For the LQ game, authors in \cite{Parise2019} show that strong monotonicty and uniform P block function properties of $F$ can represented in terms of $G$ matrix. Thus if $G$ satisfies either one of the two conditions, 
\[1-||G||_2 >0\]
\[1-||G||_{\infty}>0\]
then the social cost is continuous.



\begin{example}
We use the example in Figure \ref{fig:2}, where the condition $1-||G||_2 = 0.38 >0$ is satisfied. The social cost is observed to be continuous until the perturbations reach a network matrix where the Nash Equilibrium $\boldsymbol{x^*}$ under our assumed condition $x_i \in \mathcal{X}_i \subseteq \mathbb{R}_{\geq0}$, is not possible. The perturbations are given by $\Delta G = \begin{bmatrix} 
    0 & 0&\delta&\delta\\
    0 & 0&0&0 \\
    \delta & 0&0&0\\
    \delta & 0&0&0\\
    \end{bmatrix}$, where $\delta$ is varied as shown along x-axis in Figure \ref{fig:3}.
\begin{figure}[htbp]
\centerline{\includegraphics[width=0.7\columnwidth]{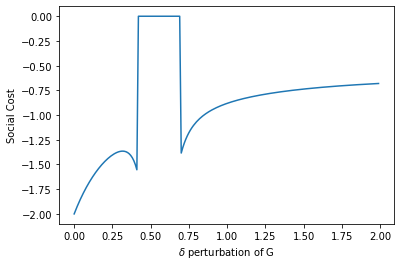}}
\caption{The social cost under perturbation of $G$ for the example in Figure 2.}
\label{fig:3}
\end{figure}

\end{example}
. 
\section{Discussion}
In this section, we discuss individual rationality constraints and analyze the applicability of such methods to large random networked agents.
\subsection{Individual rationality}
Individual Rationality is a constraint where each individual player prefers participating within the designed game to not participating. The player can expect to obtain at least as much payoff from participating than from not doing so. For the LQ network game example, we check if the cost of participating is less than that of not participating at  the Nash equilibrium and Social optimum.

Using \eqref{eq:3} in \eqref{eq:1}, we get,
\[J_i(x^*_i,z_i(x^*)) = \frac{1}{2}(x^*_i)^2 +[-x^*_i]x^*_i = -\frac{1}{2}(x^*_i)^2. \]

Thus, the cost of not participating is
\[J_i(0,z_i(x) = 0\].
Therefore, it is individually rational for the players to participate in the game.

\subsection{Large Random Networks}
A majority of  real-world networks are large networks, with numerous players and interactions. Characterizing the conditions on such networks with socially optimum Nash equilibrium, even approximately, helps identify those networks that are beneficial to society. For large network games, we assume that the adjacency matrix is a random matrix, generated through a given distribution.

Using Remark \ref{remark:1}, we know that, for a socially optimum NE in a LQ game, the adjacency matrix $G$ has to be singular. There is an opportunity here to use existing literature on random matrix theory, using standard results on its rank and invertibility to classify the class of graphs that could have socially optimum NE. For instance, the authors in \cite{Costello2008} and in \cite{Basak2018}, provide invertibility conditions on Erd\H{o}s- R\'enyi random graph, that can be used to consider graphs with socially optimum NE. As a concrete example, \cite{Costello2008} show that for $c\ln{n}/n < p < 1/2$, for $0<c<1$, and and Erd\H{o}s- R\'enyi random graph $\Gamma(n,p)$ is almost surely non-singular. This implies in the corresponding LQ game, the social optimum almost surely doesn't coincide with the NE of the game. We believe there is an immense opportunity to study the existence, stability and concurrence of NE with the social optimum for large random network games using rich literature in random matrix theory. 




\section{Conclusion}
In this paper, we present a characterization of the conditions under which networks can be designed any desired designer's objective function. We derive results for social welfare but this can applied to other instances like maximising profit, driving player interaction etc, which are dependent on player's and their neighbor's actions. We characterize conditions on the adjacency matrix for LQ games under which the Nash equilibrium matches the social optimum, thus identifying network structures which promote social good. Using a variational inequality framework, we consider uniqueness of socially optimum NE and continuity of the social cost function under perturbations of the network adjacency matrix. 

We further discuss the potential of using known properties large networks to classify the class of socially optimum NE games.
As a direction of future work, we are interested in applying the variational inequality framework to larger networks to study overall solution stability. This will enable design of networks in real world setting that improve the social welfare.
\bibliographystyle{IEEEtran}
\bibliography{library}
\end{document}